\documentclass{nature}

\usepackage{graphicx}
\makeatletter
\let\saved@includegraphics\includegraphics
\AtBeginDocument{\let\includegraphics\saved@includegraphics}
\renewenvironment*{figure}{\@float{figure}}{\end@float}
\makeatother

\usepackage{multirow}
\usepackage{mathtools}
\usepackage[colorlinks=true,linkcolor=blue,citecolor=black,urlcolor=blue]{hyperref}
\usepackage[dvipsnames]{xcolor}
\usepackage{soul}
\usepackage{authblk}
\usepackage[labelfont=bf]{caption}
\usepackage[figurename=Figure]{caption}
\usepackage{siunitx}
\usepackage{booktabs}
\usepackage{subfig}
\usepackage[T1]{fontenc}

\begin{document}

\title{Polarisation-dependent single-pulse ultrafast optical switching of an elementary ferromagnet}

\author[1,2,*]{Hanan Hamamera}

\author[3]{Filipe Souza Mendes Guimar\~aes}

\author[1]{Manuel dos Santos Dias}

\author[1,4,*]{Samir Lounis}

\affil[1]{Peter Gr\"unberg Institut and Institute for Advanced Simulation, Forschungszentrum J\"ulich \& JARA, 52425 J\"ulich, Germany}
\affil[2]{Department of Physics, RWTH Aachen University, 52056 Aachen, Germany}
\affil[3]{J\"ulich Supercomputing Center, Forschungszentrum J\"ulich \& JARA, 52425 J\"ulich, Germany}
\affil[4]{Faculty of Physics, University of Duisburg-Essen \& CENIDE, 47053 Duisburg, Germany }
\affil[*]{h.hamamera@fz-juelich.de; s.lounis@fz-juelich.de}

\maketitle


\begin{abstract}
The ultimate control of magnetic states of matter at femtosecond (or even faster) timescales defines one of the most pursued paradigm shifts for future information technology.
In this context, ultrafast laser pulses developed into extremely valuable stimuli for the all-optical magnetisation reversal in ferrimagnetic and ferromagnetic alloys and multilayers, while this remains elusive in elementary ferromagnets.
Here we demonstrate that a single laser pulse with sub-picosecond duration can lead to the reversal of the magnetisation of bulk nickel, in tandem with the expected demagnetisation. As revealed by realistic time-dependent electronic structure simulations, the central mechanism is ultrafast light-induced torques acting on the magnetisation, which are only effective if the laser pulse is circularly polarised on a plane that contains the initial orientation of the magnetisation. 
We map the laser pulse parameter space enabling the magnetisation switching and unveil rich intra-atomic orbital-dependent magnetisation dynamics featuring transient inter-orbital non-collinear states. 
Our findings open further perspectives for the efficient implementation of optically-based spintronic devices.
\end{abstract}

\maketitle 

\section*{Introduction}

The manipulation and control of magnetic materials by ultrashort laser pulses has been extensively researched since the discovery of optically-driven ultrafast demagnetisation in nickel\cite{beaurepaire1996ultrafast}.
The technological potential of this discovery was quickly recognised, leading to proof-of-concept experiments in connection with information storage\cite{stanciu2007all,lambert2014all,John2017}.
Such laser-driven magnetisation dynamics has also been explored in bulk rare-earth ferromagnets\cite{Wietstruk2011}, in ferrimagnets\cite{stanciu2007all,radu2011transient,alebrand2012interplay,Khorsand2012,Ostler2012,graves2013nanoscale,hassdenteufel2013thermally,PhysRevResearch.2.032044,van2020deterministic}, and in ferromagnetic thin films\cite{Boeglin2010,lambert2014all,vodungbo2016indirect,Vomir2017,Shokeen2017,Siegrist2019,kichin2019multiple,yamada2019efficient}.
The underlying physical picture is not yet fully understood, given the diversity of mechanisms that can contribute to or influence the dynamics on distinct or overlapping time scales~\cite{kirilyuk2010ultrafast,koopmans2010explaining,Walowski2016}.

For applications, the goal is not simply to change or demagnetise the material but to controllably reverse the magnetisation direction, which encodes an information bit.
This has been successfully achieved in GdFeCo thanks to its ferrimagnetism\cite{stanciu2007all,Ostler2012}.
For this material, the magnetisation switching is due to a laser-driven heating above the ferrimagnetic compensation point\cite{Ostler2012}, together with the different relaxation time scales of the two rare-earth and transition metal sublattices\cite{radu2011transient}, and only weakly depends on the polarisation of the laser\cite{Khorsand2012}.
Magnetisation switching has also been demonstrated for ferromagnetic Co/Pt multilayers, where the helicity of the laser is an important factor to achieve deterministic switching.
While GdFeCo can be switched with a single pulse, for Co/Pt several long pulses\cite{kichin2019multiple} or hundreds of short pulses\cite{ElHadri2016} are needed to achieve full switching --- which is detrimental for technological applications due to the high energy consumption and the relative slowness of the whole process.
It has been recently reported that only two laser pulses are enough to achieve complete helicity-dependent switching in Co/Pt\cite{yamada2019efficient}.

In order to make progress and to understand how to achieve full switching in a ferromagnet  with a single laser pulse, the appropriate physical mechanism or combination of physical mechanisms have to be identified and simulated. 
The three-temperature model\cite{beaurepaire1996ultrafast,koopmans2000ultrafast,koopmans2005unifying,koopmans2010explaining} describes the nonequilibrium thermodynamics of coupled electronic, magnetic and lattice subsystems, and provides a very good semi-phenomenological description of the demagnetisation of bulk Ni and of the switching in GdFeCo.
Demagnetisation and switching due to the stochastic magnetisation dynamics driven by laser heating of the material were studied numerically using Landau-Lifshitz-Gilbert and Landau-Lifshitz-Bloch equations\cite{radu2011transient,atxitia2011ultrafast,Mentink2012,Ostler2012}.
There are also several proposed microscopic pictures for how the electrons react to the laser and lead to ultrafast demagnetisation.
The inverse Faraday effect was proposed as a direct mechanism for laser-induced demagnetisation\cite{kimel2005ultrafast}, which evolved into a more general picture of light-induced magnetic torques\cite{tesavrova2013experimental,Berrita2016,zhang2016switching,huisman2016femtosecond,freimuth2016laser,Choi2017b}.
The superdiffusive spin transport model\cite{battiato2010superdiffusive} introduces spin-polarised hot electrons that transfer angular momentum from the magnetic atoms to a non-magnetic material.
Mechanisms for demagnetisation due to transfer of angular momentum from the spins to the lattice have also been extensively studied\cite{Stamm2007,carva2011ab,Wietstruk2011,chen2019role,maldonado2020tracking}.
Simulations considering electron-electron interactions\cite{tows2015many} identified a three-step mechanism: the laser pulse creates electron-hole excitations, spin-orbit coupling converts the excited spin to orbital angular momentum, and the latter is then quickly quenched by the lattice.
Lastly, time-dependent density functional theory simulations of the combined dynamics of the electrons and the magnetic moments that they form have yielded many microscopic insights into the ultrafast demagnetization  in the sub-\SI{100}{\femto\second} regime\cite{krieger2015laser,krieger2017ultrafast,simoni2017ultrafast,Shokeen2017,chen2019role,Siegrist2019,elliott2020microscopic}.

\begin{figure*}[ht!]
     \centering
      \includegraphics[width=1.0\textwidth]{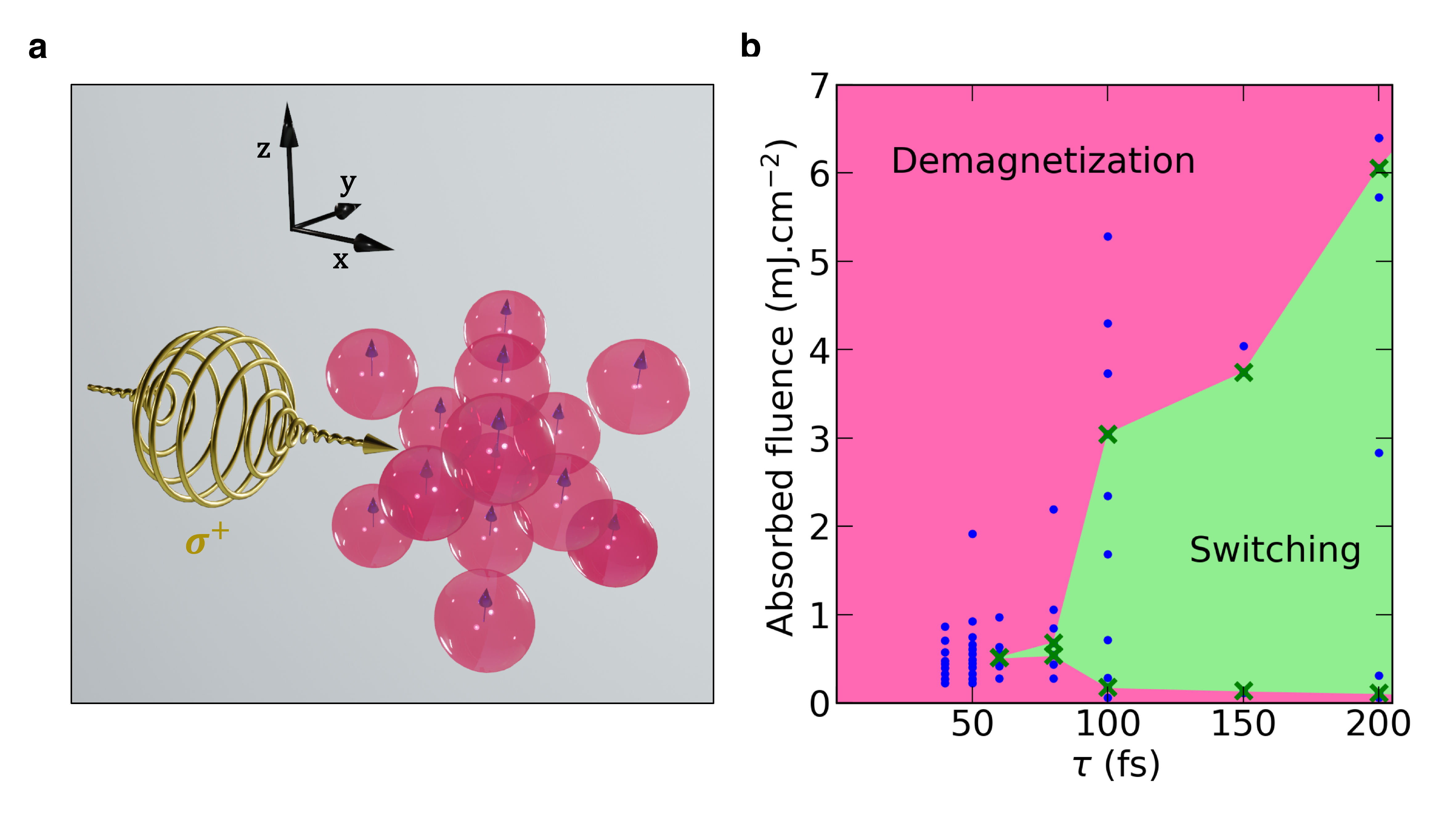}
      \caption{\small
      Parameter space for single-pulse all-optical switching of bulk Ni.
      (a) Schematic representation of a circular light pulse interacting with an fcc bulk material with initial magnetisation pointing along the $+z$ direction.
      (b) Laser parameter phase diagram for bulk Ni as a function of the laser pulse width $\tau$ and absorbed fluence.
      The dots indicate all the performed calculations, and the crosses mark the identified boundaries.
     \label{fig1} }
\end{figure*}

So far all the simulations based on a realistic description of the electronic structure were limited to ultrafast demagnetisation processes. Here we address all-optical magnetisation reversal and the possibility of inducing it with a single laser pulse in an elementary ferromagnet such as fcc bulk Ni.
We employ a recently developed time-dependent tight-binding framework parameterized from DFT calculations, with a specific algorithm enabling to monitor the non-linear magnetisation dynamics up to a few picoseconds. We show that a single laser pulse can trigger the magnetisation reversal of Ni, for which we identified the pulse parameter space enabling magnetisation switching summarized in Fig.~\ref{fig1}.
We identify ultrafast light-induced torques as the underlying mechanism, which act on the magnetisation if the polarisation of the pulse obeys specific conditions. 
We found strong non-collinear, ferromagnetic and antiferromagnetic intra-atomic transient states that are shaped by the interplay of optical inter-orbital electronic transitions and spin-orbit induced spin-flip processes.

\section*{Results}

\subsection{All-optical magnetisation reversal\label{sec3.1}}

We perform tight-binding simulations parameterized from DFT by propagating the ground state eigenvectors in real time solving the time-dependent Schrödinger equation up to a few picoseconds. 
Our Hamiltonian includes the kinetic energy as given by the electronic hopping, the spin-orbit coupling (SOC), the electron-electron exchange interactions, and the effect of the laser-induced excitations. 
The propagated solutions were then used to calculate the longitudinal and transverse components of the magnetisation as well as the absorbed energy together with the redistribution of the electronic population among the different orbitals. 
Our method enables us to use larger pulse widths to investigate the effect of both linearly and circularly polarised pulses over a long time scale (see Methods section for more details). 

In the ground state of bulk Ni, the spin moment is found to be \SI{0.51}{}$\mu_B$ and prefers to point along the cubic axes. 
Here, we assume it to point along the [001] direction, which we choose to be the $z$ cartesian axis.
We then systematically apply single optical pulses while tracking the time-dependent magnetisation dynamics of the system.
The pulses have a fixed frequency $\omega=\SI{1.55}{\electronvolt}\hbar^{-1}$ and varying widths and intensities of the electric field $E_0$, and we consider both linear and circular polarised light (see schematic Fig.~\ref{fig1}a and Methods section for more details).
 
\begin{figure*}[ht!]
    \centering
     \includegraphics[width=1.0\textwidth]{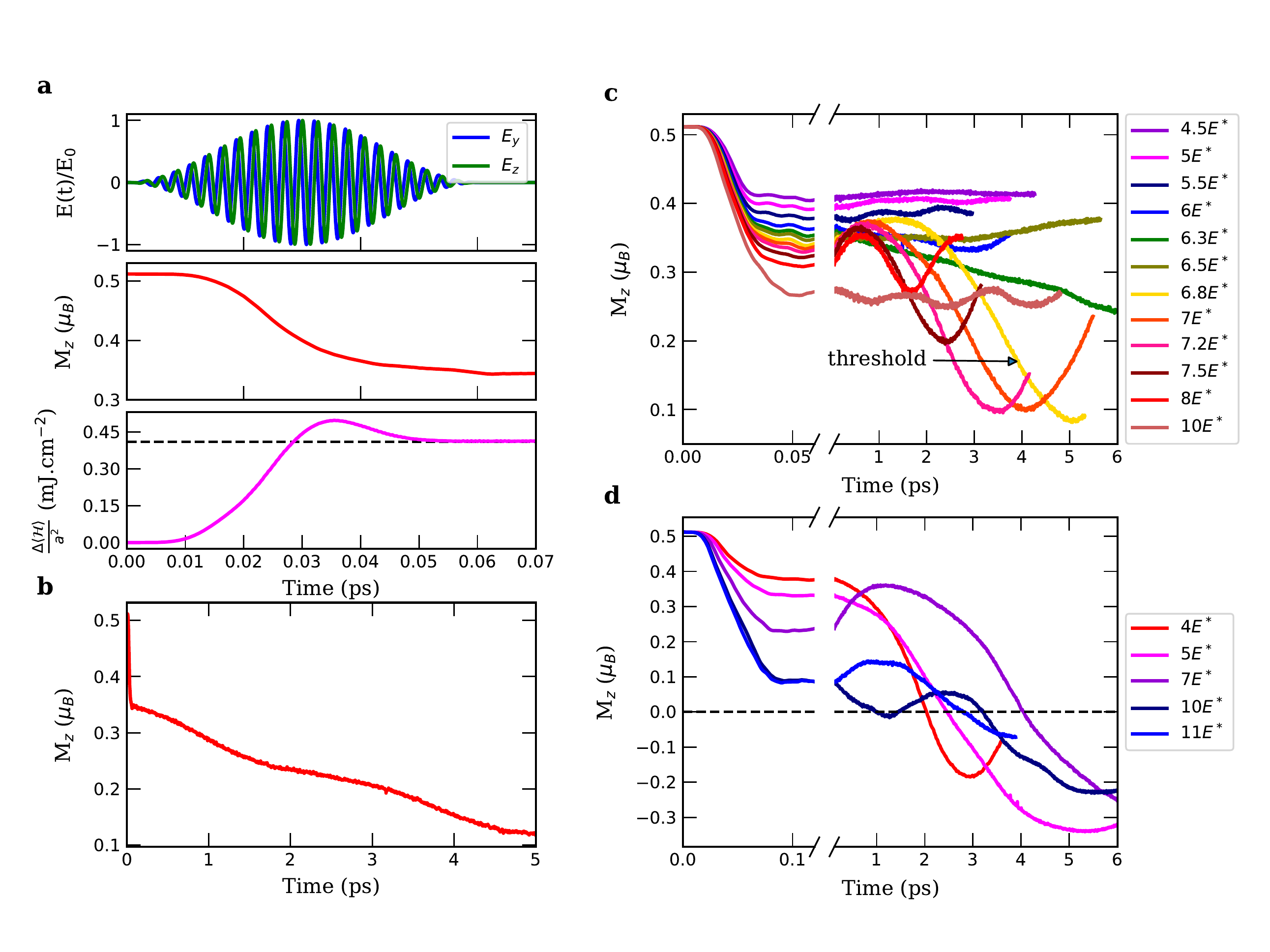}
     \caption{\small
     Diversity of laser-induced magnetisation dynamics in bulk Ni.
     (a) A \SI{60}{\femto\second}-wide pulse with intensity of $E_0 = 6E^*$, the demagnetisation that it induces, and the energy absorbed during its application.
     (b) Evolution of the $M_z$ component for longer times.
     (c) Demagnetisation curves for a \SI{50}{\femto\second}-wide pulse with different laser field intensities.
     (d) Switching is observed for a \SI{100}{\femto\second}-wide pulse for different laser field intensities.
     The reference value for the laser field intensity is $E^* = \SI{9.7e8}{\volt\meter\tothe{-1}}$.
     \label{fig2}}
\end{figure*}

As a first example, Fig.~\ref{fig2}a displays the electric field components of a \SI{60}{\femto\second} laser pulse circularly polarised in the $yz$-plane (see Methods for the precise form of the pulse), as well as the time-dependent reduction of the $z$-component of the Ni spin moment, and the absorbed laser fluence that we approximate by the change in the energy of the material divided by its cross section.
Up to \SI{70}{\femto\second}, we recover the usual demagnetisation pattern characterizing Ni. 
A reduction of about 30\% is found when the absorbed laser fluence
reaches \SI{0.41}{\milli\joule\centi\meter\tothe{-2}} at the end of the applied pulse.
This defines an initial demagnetisation region, which can be followed by a slight ``remagnetisation'' (i.e., an increase in the $z$-magnetisation), before entering another regime on longer time scales that, in this case, continues to demagnetise Ni up to \SI{5}{\pico\second}, as illustrated in Fig.~\ref{fig2}b. 
On these longer time scales one can also identify mild oscillations in the magnetisation while it continues decreasing until it reaches about 25\% of the ground state moment at \SI{5}{\pico\second}.

To characterize the effects of the laser field within the different regimes, we start by applying \SI{50}{\femto\second} pulses with different intensities. The results shown in Fig.~\ref{fig2}c exhibit a very rich scenario as a function of the reference laser electric field intensity $E^* = \SI{9.7e8}{\volt\per\meter}$.
During the initial demagnetisation regime, increasing the laser intensity leads to a stronger reduction of the spin moment, while an oscillatory behavior emerges for larger times after a clear threshold around $E_0 = 6.5E^*-6.8E^*$.
These oscillations decrease in amplitude when the intensity is further increased, and the largest responses are then limited to a relatively small range of $E_0$ values.

By increasing the laser width to \SI{100}{\femto\second}, as shown in Fig.~\ref{fig2}d, 
the laser pulse switches the sign of the $z$-component of the magnetisation. 
We also see that increasing the magnitude of the laser electric field leads to a stronger initial demagnetisation regime and also stronger oscillation amplitudes with longer periods, the latter two becoming smaller again for $E_0 = 11 E^*$.
For smaller laser field intensities, the associated switching point ($M_z = 0$) remarkably moves to earlier times.
For both \SI{50}{\femto\second}- and \SI{100}{\femto\second}-widths, the dynamical dependence on the laser intensity and width is nontrivial.
Certain combinations of intensities and widths keep the $z$-magnetisation negative for relatively long time after reversal.
Moreover, the magnetisation behavior strongly depends on the nature of the polarisation and helicity of the laser pulse (see examples in Supplementary Figure 1).
However, the oscillatory behavior and magnetisation reversal identified in panels c and d of Fig.~\ref{fig2} were only found for simulations where the laser electric field is circularly polarised and rotates in a plane that contains the initial orientation of the magnetic moment.
Therefore, unless explicitly mentioned, in the following discussions we focus on the results obtained with pulses polarised in the $yz$-plane. 

By extending the parameter space with systematic simulations of various pulses, we map all the switching and no-switching cases into the phase diagram shown in Fig.~\ref{fig1}b, where the horizontal and vertical axes represent the pulse width and the absorbed laser fluence, respectively. 
The shaded green region illustrates the switching region, where the probability of spin reversal is high. 
One can see from the diagram that there exists a critical minimum width for switching, which for fcc bulk Ni is about \SI{60}{\femto\second}.
Once the pulse width is larger than that value, we find that there is a laser fluence window for the switching to occur where the lower-bound slightly decreases and the upper-bound increases as the pulse gets wider.  
The requirement of a minimum threshold for switching is a reasonable and expected condition; however, a critical pulse energy to induce magnetisation reversal is surprising taking into consideration the demagnetization and precession induced by the laser (and illustrated in Fig.~\ref{fig2}c,d). In this sense, one would naively expect that higher energies would further excite the system leading to more spin moment loss and larger precession amplitude.
A narrow window in the same laser parameter space was experimentally identified for Co/Pt \cite{kichin2019multiple}. 
An interesting feature in the obtained demagnetisation curves is that the switching may occur while the pulse is still effective (see Supplemental Figure 2), when the system is pumped with more energetic pulses of widths larger than \SI{100}{\femto\second}. 

\subsection{Laser-induced torque\label{torque}}

\begin{figure*}[ht!]
     \centering
      \includegraphics[width=1.0\textwidth]{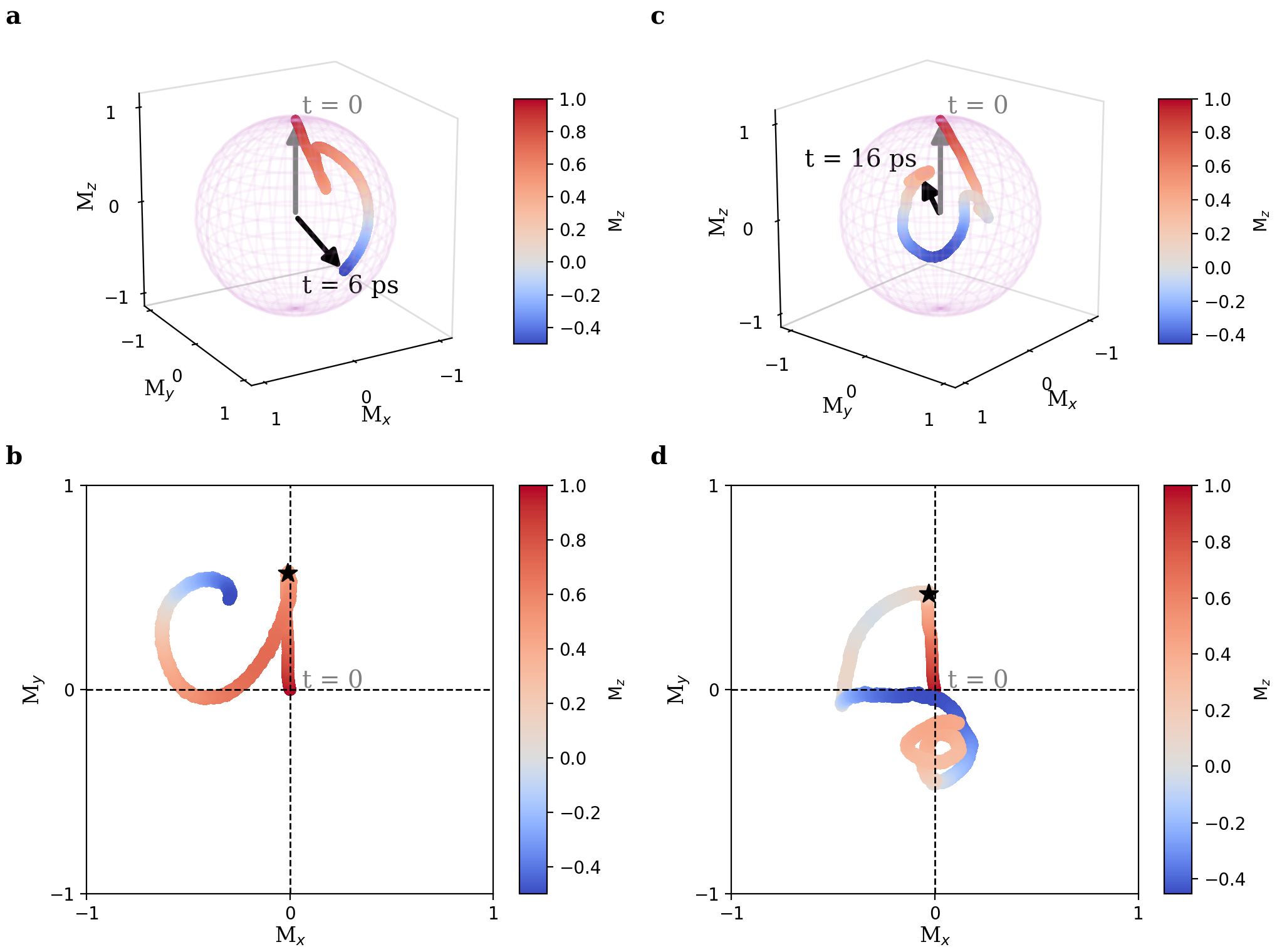}
      \caption{\small
      Three-dimensional ultrafast magnetisation dynamics of bulk Ni. (a,c) Trajectories of the magnetisation vector excited by a \SI{100}{\femto\second} pulse with intensity of (a) $E_0 = \SI{6.79e9}{\volt\per\meter}$ ($7E^*$ in Fig.\ \ref{fig2}d)
      and (c) $E_0 = \SI{9.7e9}{\volt\meter\tothe{-1}}$ ($10E^*$ in Fig.\ \ref{fig2}d), with the respective associated projection on the $xy$-plane (b,d).
      The color maps illustrate the $z$-component of the magnetisation and the star indicates the end of the pulse.
      \label{fig3}}
\end{figure*}

The dynamics of the magnetisation in each scenario is actually more complicated than its $z$-component in Fig.~\ref{fig2} can display---the transverse components also change dramatically depending on the laser characteristics.
In Fig.~\ref{fig3}, we show the 3-dimensional (3D) magnetisation response to a \SI{100}{\femto\second} laser pulse with two different intensities.
The lower intensity case in Fig.~\ref{fig3}a illustrates the magnetisation quickly reducing in amplitude and early on rotating in the $yz$-plane (as seen also from the transverse cut shown in Fig.~\ref{fig3}b)---interestingly, the plane containing the initial state and the polarisation of the laser pulse.
After the fast initial demagnetisation region in the $yz$-plane, the magnetisation acquires also an $x$-component and rotates away from the initial plane. 
When increasing the laser intensity, Figs.~\ref{fig3}(c-d), the magnetisation can even switch back towards the initial direction of the moment after some time. 

The identified magnetisation precession results from a torque induced by the laser. 
As we consider a bulk material and no applied magnetic field, the only torque acting on the spin magnetisation $\mathbf{M} = -\gamma \langle\mathbf{S}\rangle$ is the spin-orbit torque $\mathbf{T}_\text{SOC} = -\gamma \lambda \langle\mathbf{L}\times \mathbf{S}\rangle$.
Here $\gamma$ is the gyromagnetic ratio, $\lambda$ is the spin-orbit interaction strength, and $\mathbf{S}$ and $\mathbf{L}$ are the quantum-mechanical spin and orbital angular momentum operators, respectively.
The expectation value is computed with the time-dependent electronic wave functions.
The magnetisation dynamics due to the combination of precession and demagnetisation can then be written as
\begin{equation}\label{eq:torques}
  \frac{\mathrm{d}\mathbf{M}}{\mathrm{d}t} = \mathbf{T}_\text{SOC}
  = -\gamma \mathbf{M} \times \left(\mathbf{B}_\mathrm{IFE} + \mathbf{B}_\mathrm{MAE}\right) - \chi_\mathrm{L} \mathbf{M} \;.
\end{equation}
The effective field that generates the torque has two distinct contributions: $\mathbf{B}_\mathrm{IFE}$ is due to an inverse Faraday-like effect (IFE) and acts only during the laser pulse, while $\mathbf{B}_\mathrm{MAE}$ is due to the time-dependent magnetic anisotropy of the non-equilibrium electronic system and so also acts after the laser pulse is over.
The last term is a longitudinal contribution that represents the demagnetisation driven by the laser with rate $\chi_\mathrm{L}$ and is also contained in $\mathbf{T}_\text{SOC}$.

\begin{figure*}[ht!] 
    \centering
     \includegraphics[width=0.9\textwidth]{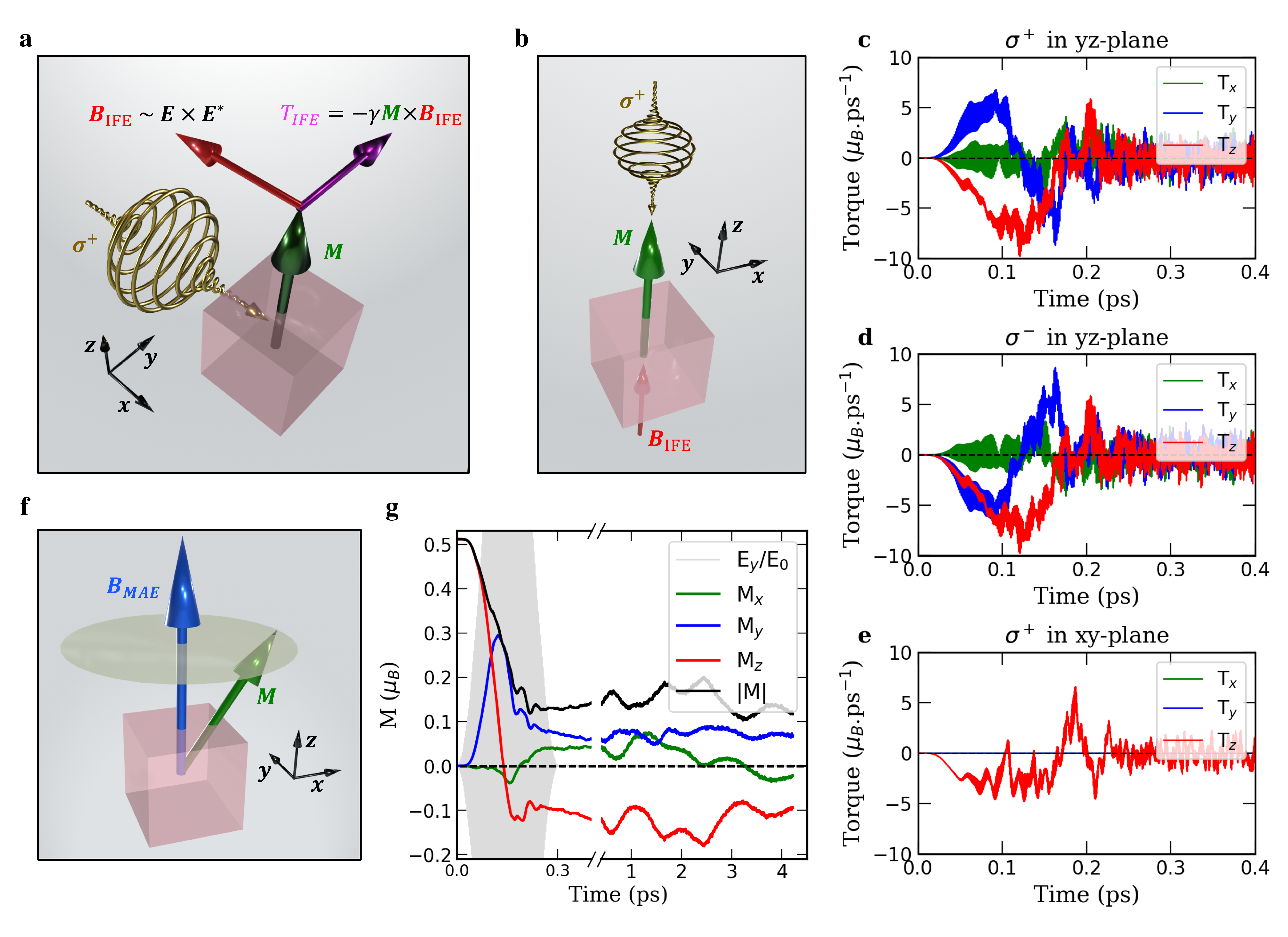}
     \vspace{-2em}
     \caption{\small
     Laser-induced ultrafast magnetic torques on bulk Ni.
     (a,b) Schematic illustrations representing a laser acting on an initial magnetisation along the $z$-axis: (a) a right-circularly polarised pulse ($\sigma^+$) in the $yz$-plane induces an inverse Faraday-like magnetic field opposite to the polarisation direction ($-x$), which at $t=0$ exerts a torque on the magnetisation along the $+y$ direction;
     (b) a right-circularly polarised pulse in the $xy$-plane would generate an inverse Faraday-like magnetic field parallel to the magnetisation ($+z$) and so at $t=0$ it exerts no torque.
     (c,d,e)
     The components of the torques induced by a \SI{300}{\femto\second} wide pulse with intensity of $E_0 = \SI{9.7e9}{\volt\per\meter}$ that is (c) right-circularly polarised in the $yz$-plane, (d) left-circularly polarised ($\sigma^-$) in the $yz$-plane, and (e) right-circularly polarised in the $xy$-plane, respectively, computed according to Eq.~\eqref{eq:torques}.
     (f) The magnetisation dynamics also has a contribution from the internal field due to the magnetic anisotropy which induces precession.
     (g) The three components of magnetisation along with the total length, corresponding to the case shown in (c).
    \label{fig4} }
\end{figure*}

We now consider a \SI{300}{\femto\second} laser pulse, in order to enhance the torque contribution which is driven directly by the laser. 
$\mathbf{B}_\mathrm{IFE}$ is expected to point perpendicular to the polarisation plane of the circular laser pulse, i.e. along the $x$-direction, given that the electric field rotates in the $yz$-plane as shown in Fig.~\ref{fig4}a.
The resulting torque points along $y$ and enforces the observed rotation within the $yz$-plane.
Inverting the polarisation of the laser pulse changes the direction of the torque and so also the sense of rotation of the magnetisation.
This can be identified in Fig.~\ref{fig4}(c-d), where the total torque is plotted as function of time.
At early times, one notices that the three components of the torques are finite. The $y$-component of the torque acts earlier and is larger than the $x$-component. Both components change sign by switching the polarisation of the pulse.
If the polarisation plane is perpendicular to the ground state magnetisation, as illustrated in Fig.~\ref{fig4}b, the corresponding transverse torque cancels out (see Fig.~\ref{fig4}e), and no rotation of the magnetisation is expected as verified in our simulations.
Therefore, a laser-driven torque leading to magnetisation reversal is most efficient if the initial magnetisation direction is contained in the plane of the circular polarisation.
After the laser pulse is over, the second torque shown in Eq.~\eqref{eq:torques} kicks in and rotates the moment out of the $yz$-plane (sketched in Fig.~\ref{fig4}f) at a time scale of the order of picoseconds, settled by the non-equilibrium magnetic anisotropy energy.
The magnetisation dynamics on this longer time scale is shown in Fig.~\ref{fig4}g, where we can also identify oscillations in the magnitude of the magnetisation.
These hint at internal dynamics that we now discuss.

\subsection{Orbital-dependent magnetisation dynamics.\label{orbital}}

\begin{figure}[!htp]
  \subfloat{\label{fig1:a}\includegraphics[width=0.58\linewidth]{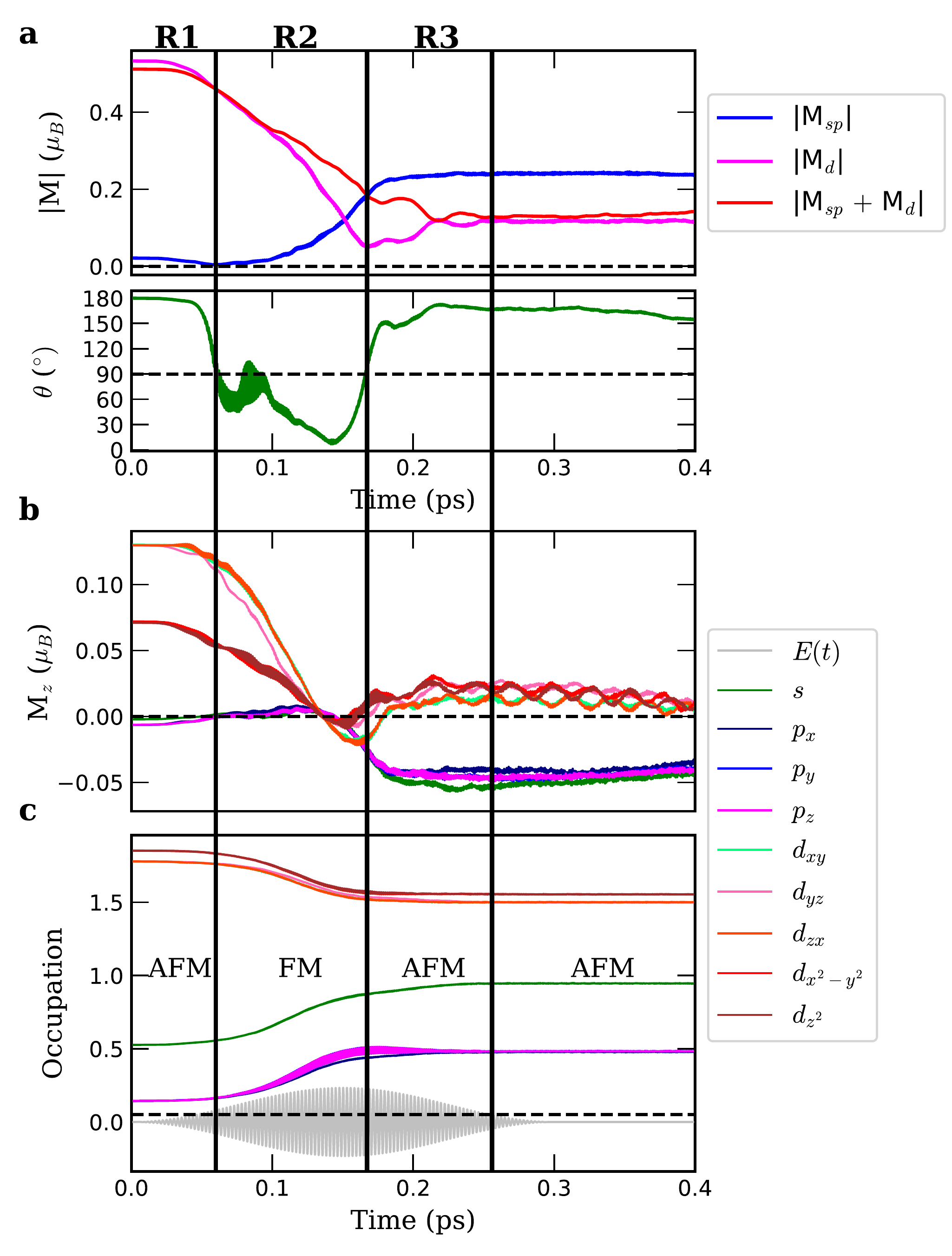}}\hfill
  \subfloat{\label{fig1:b}\includegraphics[width=0.41\linewidth]{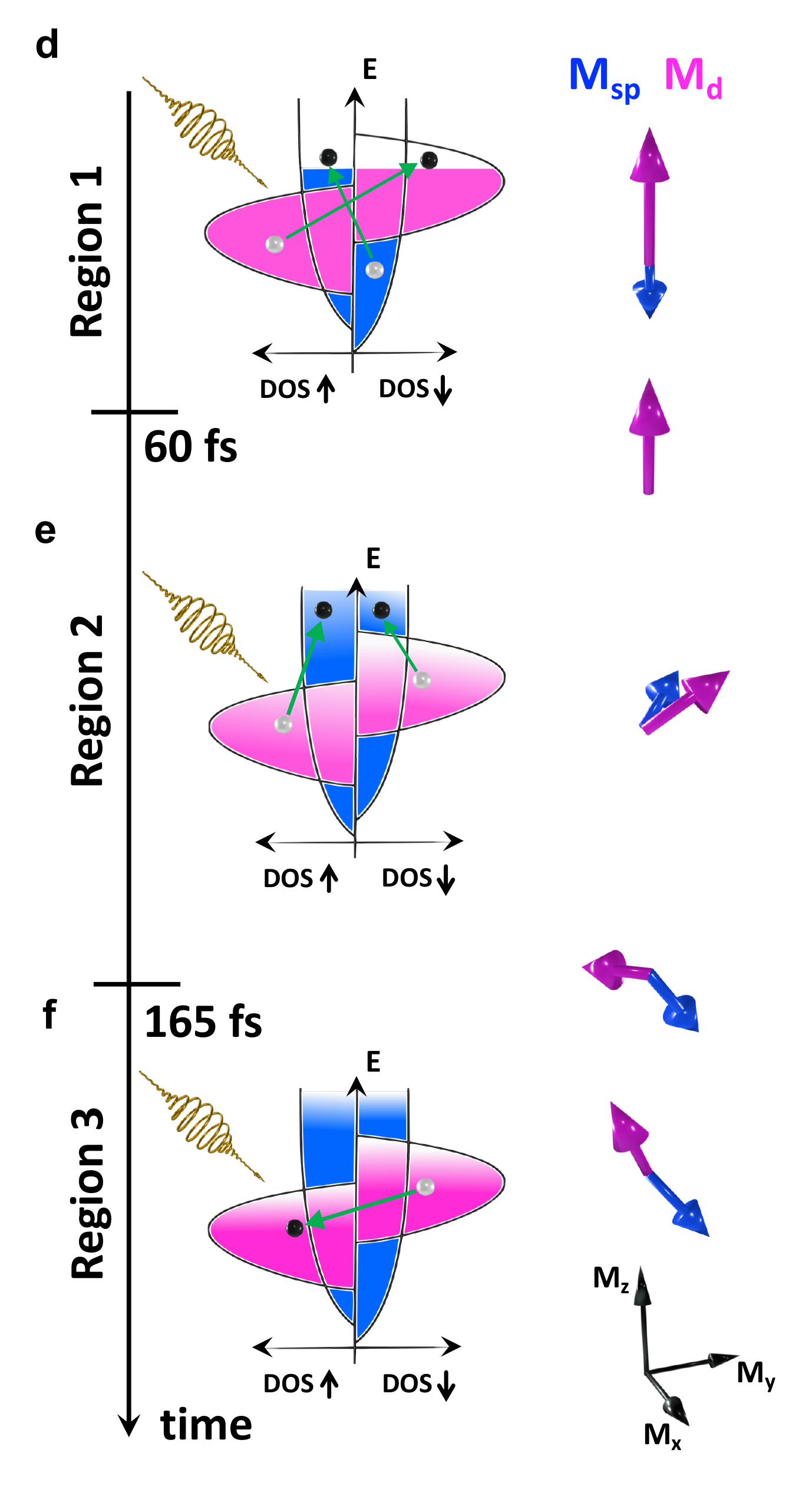}}
  \caption{\label{fig5} \small
  Orbital dependence of the ultrafast magnetisation dynamics of bulk Ni.
      (a) The length of the magnetisation vectors for the sp and d orbitals along with their vector sum and the angle $\theta$ between the magnetisation vectors of the sp and d orbitals.
      (b) The orbital contributions to the $z$ component of the magnetisation.
      (c) The occupation of the sp and d orbitals.
      The vertical lines indicate the identified three regimes of demagnetisation.
      (d-f) Schematics of the magnetisation dynamics for the different regimes.
      The diagrams on the left column indicate the main electronic processes while those on the right illustrate the evolution of the noncollinear intra-atomic magnetic moments.
      In this simulation the width of the laser pulse was \SI{300}{\femto\second} and its intensity $E_0 = \SI{9.7e9}{\volt\per\meter}$.}
\end{figure}

Our method enables us to study not only the time-dependent amplitude of the total magnetisation but also the internal dynamic contributions from different electronic states, as we now discuss for the same circular right-handed pulse of \SI{300}{\femto\second} width polarised in the $yz$ plane.
We compute the magnetisation vectors $\mathbf{M}_\mathrm{sp}$ and $\mathbf{M}_\mathrm{d}$ carried by the sp and d orbitals, respectively, with their magnitudes and the angle between them shown in Fig.~\ref{fig5}a.
The dynamics of the $z$ component of the magnetisation contributed by each individual orbital is given in Fig.~\ref{fig5}b and their occupations in Fig.~\ref{fig5}c, showing the complex and rich behavior originating from each s, p and d orbital to the dynamics up to \SI{0.4}{\pico\second}.
We clearly observe that the sp orbitals contribute to the magnetisation dynamics on an equal footing and undergo switching similarly to the d orbitals.
The most striking finding is of a transient intra-atomic non-collinear state of the sp and d magnetisations, with an effective ferromagnetic coupling lasting from about \SI{60}{} to \SI{165}{\femto\second},
which is reminiscent of the transient ferromagnetic state of the two sublattices in ferrimagnetic alloys undergoing ultrafast magnetisation switching\cite{radu2011transient}.
After \SI{165}{\femto\second}, the coupling in the transient non-collinear state switches to antiferromagnetic, which settles in an antiferromagnetic state with weak intra-atomic non-collinearity once the laser weakens and after it ends.

We can identify three main regions with distinct dynamical behavior while the laser is acting on the material.
In the first region, up to about \SI{60}{\femto\second}, there is a small reduction of the magnetisation of the d states, while the one contributed by the sp states falls to zero at the end of this region.
The dynamics of all the orbitals belonging to the respective s, p and d groups closely follow each other, with the $d_{yz}$ orbital starting to split from the other d orbitals (Fig.~\ref{fig5}b), while their occupations change little (Fig.~\ref{fig5}c).
As is well known, the sp and d orbitals are antiferromagnetically coupled to each other in the ground state, but surprisingly the transition to the next dynamical region is accompanied by a strong noncollinearity of $\mathbf{M}_\mathrm{sp}$ and $\mathbf{M}_\mathrm{d}$ (Fig.~\ref{fig5}a).
In the second dynamical region, from about \SI{60}{} to \SI{165}{\femto\second}, there is a strong collapse of the magnetisation of the d states accompanied by a strong increase of the magnetisation of the sp states (Fig.~\ref{fig5}a).
The angle between the two magnetisation vectors varies in a complex way and their coupling becomes ferromagnetic-like, with an accompanying rotational motion of the total magnetisation vector in the $yz$-plane.
The switching of the $z$ component of the magnetisation occurs due to the large transfer of spin angular momentum from the d to the sp states (Fig.~\ref{fig5}a), which is now also accompanied by a large transfer of orbital population (Fig.~\ref{fig5}c).
At the end of this region, the $d$ magnetisation is minimal and is in the process of rotating from being parallel to being antiparallel to the larger magnetisation now displayed by the $sp$ states.
In the third dynamical region, from about \SI{165}{\femto\second} to essentially the end of the laser pulse, the $d$ magnetisation partly recovers and assumes an almost antiparallel alignment to the $sp$ one (Fig.~\ref{fig5}a).
The orbital occupations stabilize (Fig.~\ref{fig5}c), but the $d$ orbitals develop internal oscillations with a short period of tens of femtoseconds, which continue after the laser is over (Fig.~\ref{fig5}b).

The previous observations lead us to propose the following physical picture for the different dynamical processes actively driven by the laser pulse, as illustrated in Fig.~\ref{fig5}d-f. 
In the first demagnetisation region up to \SI{60}{\femto\second}, intra-orbital spin-flip processes, i.e.\ within each orbital channel (d-d), (sp-sp) are responsible for the initial reduction of the magnetisation until the spin moment of the sp-electrons is fully quenched (Fig.~\ref{fig5}d).
Both the mechanism and the time scale are explained by spin-orbit coupling. 
In the second region starting after \SI{60}{\femto\second}, inter-orbital optical transitions become important while maintaining strong intra-atomic noncollinearity with the sp and d moments entering a transient ferromagnetic coupling and reaching similar magnitudes (Fig.~\ref{fig5}e).
Here the time scale is set by effective inter-orbital exchange interactions.
The nature of the effective inter-orbital exchange coupling changes due to the orbital repopulation.
After \SI{165}{\femto\second}, a new equilibrium between the occupations of the orbitals is reached, which recovers the initial inter-orbital antiferromagnetic coupling, enforcing the weaker moment (which is now the d-magnetisation) to point in the opposite direction to the stronger one originating from the sp states (Fig.~\ref{fig5}g).
There is also a significant remagnetisation of the d orbitals which could be driven by the coupling to the larger magnetisation of the sp orbitals and is assisted by the laser (Fig.~\ref{fig5}a).
To summarize, the demagnetisation rate of both types of electrons is not the same since the strength of the matrix elements responsible for the spin-orbit driven spin-flip processes is orbital dependent.
When both families of orbitals are demagnetized, the strong population switching of the electronic states in favor of the sp-type forces the d-electrons to have their moments growing in the direction opposite than that of the sp-electrons when their natural inter-orbital antiferromagnetic coupling is restored. 

\section*{Discussion}
In conclusion, we predict via time-dependent electronic structure simulations that the so far elusive magnetisation switching in an elementary ferromagnet such as bulk fcc Ni is possible with a single laser pulse.
We mapped the laser-pulse parameter regime enabling the reversal of the magnetisation and found that a minimum pulse width of \SI{60}{\femto\second} is required, while increasing the pulse width widens the laser fluence range allowing all-optical manipulation of the direction of the magnetisation. 
The magnetisation reversal is enabled by laser-driven torques that rotate its orientation together with the usual demagnetisation process, and require the proper selection of a laser pulse which is circularly polarised in a plane containing the ground state magnetisation.
Our simulations unveiled complex and rich orbital magnetisation dynamics with transient intra-atomic non-collinear states and unexpectedly fast precessional dynamics which we attribute to the nonequilibrium magnetic anisotropy created by the laser-induced electronic repopulation.
 Even though relaxation mechanisms were not incorporated in our simulations, we conjecture that adding a dissipation channel (for instance due to electron-phonon coupling) would both enhance the computed demagnetisation and dampen the laser-driven precession without 
 qualitatively affecting our main findings, i.e.\ the presence of complex intra-atomic spin dynamics together with the light-induced torque, which if correctly positioned can efficiently switch the magnetisation of an elementary ferromagnet.
We envision that our results will promote further studies focusing on the polarisation-dependent all-optical magnetisation reversal and even in characterizing intriguing inter-orbital intra-atomic transient complex magnetic states.
Such findings open further perspectives in the implementation of all-optical addressed spintronic storage and memory devices.

\begin{methods}

\subsection{Theory}\label{sec4.1}
We utilize a multi-orbital tight-binding Hamiltonian that takes into account the electron-electron interaction through a Hubbard like term and the spin-orbit interaction, as implemented on the TITAN code to investigate dynamics of transport and angular momentum properties in nanostructures~\cite{guimaraes2015fmafm,guimaraes2017dynamical,guimaraes2019comparative,guimaraes2020spin}. 
To describe the interaction of a laser pulse with the system, we include a time-dependent electric field described by a vector potential $\mathbf{A}(t) = -\int \mathbf{E}(t)dt $. The full Hamiltonian is given by
\begin{equation}
\mathcal{H}(t)= \mathcal{H}_\mathrm{kin} + \mathcal{H}_\mathrm{xc} + \mathcal{H}_\mathrm{soc} - \int\mathrm{d}\mathbf{r}\; \hat{\mathbf{J}}^\text{C}(\mathbf{r},t)\cdot \mathbf{A}(\mathbf{r},t) \;.
\end{equation}
More details on each term can be found in Supplementary Note 1.
The dipole approximation was used in the implementation of the vector potential, meaning that the spatial dependency is not included since the wavelength of the used light ($\hbar\omega = \SI{1.55}{\electronvolt} \rightarrow \lambda = \SI{800}{\nano\meter}$) is much larger than the lattice constant, and that the quadratic term as well as the other higher terms are zero~\cite{chen2019revealing}.
We approximate the absorbed laser fluence by the change in the energy of the system divided by its cross section,
\begin{equation}
    \text{absorbed laser fluence} = \frac{\langle\mathcal{H}\rangle(t) - \langle\mathcal{H}\rangle(0)}{a^2} \;,
\end{equation}
where $a$ is the lattice constant.
This approaches a stable value after the end of the laser pulse.

\subsection{Pulse shape} \label{sec4.2}
For the right-handed circular pulse ($\sigma^+$) polarised in the $yz$-plane, for example, the pulse shape is described  using a vector potential of the following form~\cite{volkov2019attosecond},
\begin{equation}\label{eq:A(t)_circular}
\mathbf{A}(t) = -\frac{E_0}{\omega} \cos^2({\pi t}/{\tau}) [\sin(\omega t) \mathbf{\hat{y}} - \cos(\omega t) \mathbf{\hat{z}}],
\end{equation}
where $E_0$ is the electric field intensity, $\tau$ is the pulse width, $\omega$ is the laser central frequency which is set to $\SI{1.55}{\electronvolt}\hbar^{-1}$.
The magnetic field of the laser is neglected since it is much smaller than the electric field.

For the linear pulse ($\pi$) of a propagation direction along the diretion $\mathbf{\hat{u}}$, using the same central frequency, the vector potential is described as
\begin{equation}\label{eq:A(t)_linear}
\mathbf{A}(t) = -\frac{E_0}{\omega} \cos^2({\pi t}/{\tau}) \sin(\omega t) \mathbf{\hat{u}} \,.
\end{equation}

\subsection{Computational details}
Calculations were performed on bulk face centered cubic Nickel using the theoretical lattice constant of \SI{3.46}{\angstrom} given in Ref.\citenum{Papaconstantopoulos:2015hr}, one atom in the unit cell for all calculations, a uniform k-point grid of ($22\times22\times22$) and a temperature of \SI{496}{\kelvin} in the Fermi-Dirac distribution.
The initial step size for the time propagation is $\Delta t = 1$ a.u.\ which changes in the subsequent steps to a new predicted value such that a relative and an absolute error in the calculated wave functions stay smaller than $10^{-3}$~ \cite{Hairer2010}.

We tested the results for accuracy by increasing the number of k-points and decreasing the tolerance for the relative and absolute errors.
The method was also tested for stability by changing one of the laser parameters by a very small number while keeping the other parameters fixed, for one case that we already have results for.
The results then were not very different~\cite{higham2002accuracy}. 

\end{methods}

\noindent \textbf{Code availability} The tight-binding code that supports the findings of this study, TITAN, is available from the corresponding author on request.

\noindent \textbf{Data availability} The data that support the findings of this study are available from the corresponding authors on request.

\begin{addendum}

\item This work was supported by the Palestinian-German Science Bridge BMBF program and the European Research Council (ERC) under the European Union's Horizon 2020 research and innovation programme (ERC-consolidator Grant No. 681405-DYNASORE).
The authors gratefully acknowledge the computing time granted through JARA on the supercomputer JURECA\cite{jureca} at Forschungszentrum Jülich.

\item[Author contributions] H.H. performed the calculations and implemented the time evolution method under the supervision of F.S.M.G and M.d.S.D. S.L. initiated, designed and supervised the project.
All the authors discussed the obtained results and contributed to writing and revising the manuscript.

\item[Competing Interests] The authors declare no competing interests.

\end{addendum}

\section*{References}
\bibliographystyle{naturemag}

\end{document}